# Unveiling the Role of 2D Monolayer Mn-doped MoS$_2$ Material: Toward an Efficient Electrocatalyst for H$_2$ Evolution Reaction

Joy Ekka[1], Shrish Nath Upadhyay[2], Frerich J. Keil[3] and Srimanta Pakhira[1, 2, 4*]

[1] Department of Physics, Indian Institute of Technology Indore (IIT Indore), Simrol, Khandwa Road, Indore-453552, MP, India.
[2] Department of Metallurgy Engineering and Materials Science (MEMS), Indian Institute of Technology Indore (IIT Indore), Khandwa Road, Simrol, Indore-453552, MP, India.
[3] Department of Chemical Reaction Engineering, Hamburg University of Technology, 21073 Hamburg, Germany.
[4] Centre for Advanced Electronics (CAE), Indian Institute of Technology Indore (IIT Indore), Khandwa Road, Simrol, Indore-453552, MP, India.

*Corresponding author: spakhira@iiti.ac.in (or) spakhirafsu@gmail.com

# Abstract:

Two-dimensional (2D) monolayer pristine MoS$_2$ transition metal dichalcogenide (TMD) is the most studied material because of its promising aspects as nonprecious electrocatalyst for hydrogen evolution reaction (HER). Previous studies have shown that the basal planes of the 2D MoS$_2$ are catalytically inert and hence, they cannot be used directly in desired applications such as electrochemical HER in industries. Here, we have thoroughly studied the defect-engineered Mn-doped 2D monolayer MoS$_2$ (Mn-MoS$_2$) material where Mn was doped in the pristine MoS$_2$ to activate the inert basal planes. Using density functional theory (DFT) method, we performed rigorous inspection of electronic structures and properties of the 2D monolayer Mn-MoS$_2$ to be a promising alternative to noble metal free catalysts for the effective HER. Periodic 2D slab of the monolayer Mn-MoS$_2$ was created to study the electronic properties (such as band gap, band structures and total density of states (DOS)) and the reaction pathways occurring on the surface of the material. The detailed HER mechanism has been explored by creating the Mn$_1$Mo$_9$S$_{21}$ non-periodic finite molecular cluster model system using M06-L DFT method including solvation effects to determine the reaction barriers and kinetics. Our study reveals that the 2D Mn-MoS$_2$ follows the most favorable Volmer-Heyrovsky reaction mechanism with very low energy barriers during the H$_2$ evolution. It was found that the change of free energy barrier (ΔG) during the Heyrovsky reaction is about 10.34 - 10.79 kcal mol$^{-1}$ (computed in solvent phase), indicating an exceptional electrocatalyst for HER. The Tafel slope is lower in the case of 2D monolayer Mn-MoS$_2$ material due to the overlap of the *s*-orbital of the hydrogen and *d*-orbitals of the Mn atoms appeared in the HOMO and LUMO transition states (TS1 and TS2) of both the Volmer and Heyrovsky reaction steps. The better stabilization of the atomic orbitals in the HER rate-limiting step Heyrovsky TS2 is a key for reducing the reaction barrier, thus the overall catalysis indicating a better electrocatalytic performance for H$_2$ evolution. This study is focused on designing low cost and efficient electrocatalysts for HER by using earth abundant transition metal dichalcogenides (TMDs) and decreasing the activation energy barriers by scrutinizing the kinetics of the reaction for reactivity.

# 1 Introduction:

The world's energy supply is heavily dependent on fossil fuels and energy production emanating from this source releases substances that are harmful for the environment. Rapid depletion of this source is also another concern which drives the necessity to find eco-friendly and zero emission energy sources. To resolve this, we need to increase the usage of renewable energy sources to meet future energy demands. Hydrogen offers itself as a new and unconventional renewable wellspring due to its high energy density and zero emission greenhouse gases among other available alternative energy sources or fuels.[1] To produce clean and efficient hydrogen, electrolysis of water has been a viable way ever since its discovery in 1789.[2] Electrochemical water splitting is a sustainable strategy to produce hydrogen (and oxygen) and replace conventional fossil fuels. Hydrogen is a non-polluting energy source as water itself is a product of H$_2$ combustion. However, the conventional water splitting process cannot just separate water into hydrogen and oxygen, and just like any other chemical processes even this reaction needs energy input to overcome the barrier in the electrochemical process. Thus, the electrochemical water splitting requires highly active catalysts to bring down the overpotential needed to produce hydrogen (+1.23V; 25°C; 1 atm). Hydrogen evolution reaction (HER) i.e., simply 2H$^+$ + 2e$^-$ → H$_2$ is a multiphase reaction for the sustainable production of hydrogen ($H_2$) and can occur via two processes, either through Volmer-Heyrovsky process or Volmer-Tafel process.[3–5]

However, the mentioned reactions are competitive processes, and they are dependent on the electronic structure of the electrode surface. Boosting the efficiency of sluggish HER process is a challenging task at hand. As of now Platinum (Pt) is the best-known electrocatalyst for HER due to its zero overpotential in acidic electrolytes.[6] Because of the optimum Gibbs free energy (G) for adsorption of atomic hydrogen, binding energy and desorption of hydrogen from the surface with low activation energies, Pt based catalysts have been long known as effective HER electrocatalysts.[7,8] However, the limited availability and high cost of Platinum inhibit its usage as an efficient catalyst at large scale in commercial and industrial applications. To produce hydrogen on a global scale, we need to cut down the cost of production. Reducing the dependency on noble metal-based catalysts or to reinstate them even completely with non-noble metal alternatives would be a driving step towards hydrogen economy. So, the urge for search of Pt free catalyst for HER is of paramount importance in modern materials science and technology.

From past few decades countless exhaustive research works have been focused on the development of earth abundant and ecofriendly materials showing excellent electrocatalytic effect for the efficient production of $H_2$ through HER. Many of them such as Pt, Au, and Pd noble metal based electrocatalysts, nano-porous materials, two-dimensional transition metal dichalcogenides (2D TMDs), many 2D-3D material alloys and 2D TMDs doped with other atom species have a significant impact as an efficient HER catalyst.[9–18] A plethora of breakthroughs has been achieved in this regard for rational design of HER electrocatalysts. In recent times, earth abundant layered TMDs, for example $MoS_2$, have attracted tremendous interest due to its inherent properties which produce hydrogen at very low overpotential with high current density.[15,19,20] This is greatly recognized that the 2D monolayer $MoS_2$ material can exist in several possible structures such as 1T (octahedral structure), 1T$^{'}$ (distorted octahedral structure) and 2H (hexagonal structure).[21] The 2H phase structure (i.e., 2D monolayer 2H-$MoS_2$) is the most stable[21] and they have been commonly used for HER electrocatalysis.[22–24] The high activity of 2D monolayer $MoS_2$ is because of its appropriate Gibbs free energy of adsorbed atomic hydrogen.[25] Other $MoS_2$ based materials such as $MoS_2$ di-anionic surface with controlled molecular substitution of S sites by –OH functional group have also proven to be efficient electrocatalysts.[26] Although this material shows promising aspects, but it is still not sufficient in its present form for large industrial and commercial application purposes due to its inert basal plane. Theoretical calculations indicated that the P-doped MoS$_2$ shows a good catalytic activity for HER by reducing the change of free energy (ΔG). This is due to the P-doping in the pristine MoS$_2$ which activated the inert basal plane of it,[27] but, this non-metal doping is quite difficult because of instantaneous formation of MoP.[10] To avoid formation of MoP, quite expensive apparatus such as plasma ion implantation are required and, in some cases, especial precautions have to be taken during the P-doping in the MoS$_2$.[28] Thus, another method or material is required for developing low cost and efficient HER catalysts to produce desirable H$_2$ for industrial and commercial applications. Several techniques have been developed to generate high-performance TMD-based materials, such as defect engineering, metal-atom doping, nanostructure engineering, interface and strain engineering, and phase engineering.[15,29] Another problem frequently appears in the 2D TMDs due to the stacking feature of the 2D layers that decreases the number of exposed sites and the conductivity along two stacked layers is extremely low,[30] thereby impeding charge transfer and decreasing the HER performance of the 2D TMDs. A prominent factor controlling the rate of HER is vested upon the fact that the pristine 2D monolayer MoS$_2$ shows semiconducting properties indicating low conductivity for electrons, thus being inadequate for large commercial applications. One of the promising ways for enhancing HER is to expose the active sites of the pristine 2D monolayer 2H-MoS$_2$.[31] It was found that the most of the active sites of the pristine 2D monolayer 2H-MoS$_2$ for HER are located at the Mo and S edge sites.[32] In order to modulate the electron transport for achieving proper conducting pathway and enhance the hydrogen evolution, the doping of external elements in the pristine 2D monolayer 2H-MoS$_2$ nanostructure appropriately is the promising way in the modern technology.[14] Therefore, the mechanistic insights are relevantly of paramount importance while designing efficient electrocatalysts for H$_2$ evolution.

The development of operative, stable, and economic HER catalyst to overcome the challenges associated with H$_2$ production from electrolysis of water is a salient comprehension for driving down the production cost and extension of hydrogen economy. For example, engineering the HER activity of MoS$_2$ via co-confining selenium and cobalt in the surface and inner plane respectively has shown promising results.[33] We have proposed that Mn-doping in the pristine 2D monolayer MoS$_2$ TMD material can activate the inert basal plane and the Mn-doped 2D MoS$_2$ (in short Mn-MoS$_2$) can be a promising material for an efficient H$_2$ evolution.[34,35] For the transition metal-based catalysts, their performance is correlated to their surface electronic structures and the electronic configuration of the *d*-orbital of the transition metal.[11] In this regard, we have computationally developed two dimensional (2D) single layer Mn doped $MoS_2$ material (i.e. Mn-$MoS_2$) and investigated its electrocatalytic performance for efficient HER. First, we performed the first principles-based quantum mechanical (QM) hybrid periodic density functional theory (DFT)[36–41] calculations to obtain the electronic properties like electronic band structures, band gap and total density of states (DOS). Recently, QM DFT approaches, and molecular simulations have been employed for modeling heterogeneous catalytic reactions, adsorption, and chemical reactions on the 2D metal surfaces.[42–45] We found out that the 2D monolayer Mn-$MoS_2$ shows zero band gap due to Mn-doped in the pristine 2D $MoS_2$. The density of states calculation indicates that there is a large number of electronic states available around the Fermi energy ($E_F$) level with a high availability of electrons due to the doping of Mn atoms in the pristine 2D monolayer $MoS_2$ material.

One of the key features in determining the smooth flow of reaction is the change in free energy (ΔG) of the possible reaction intermediates. So, to screen an appropriate candidate among the options available, it is important to compute the value of ΔG during hydrogen adsorption and this is an important parameter for evaluating the catalytic activity during HER process. Lately the quantum computational method has provided practicable procedures for calculating the free energy changes based on the density functional theory (DFT).[46–49] By modeling the possible reaction intermediates during the hydrogen evolution process on the surface of the electrocatalyst, thermodynamical properties can be obtained by using DFT methods. Therefore, we prepared a finite molecular cluster model system of the Mn-$MoS_2$ material and carefully studied each and every reaction intermediate appeared during the HER process by employing DFT method in both gas and solvent phase calculations. Our study showed that the 2D monolayer Mn-$MoS_2$ TMD shows an excellent catalytic activity for $H_2$ evolution.

# 2 Methodology and Computational Details:

**2.1 Periodic Structure Calculations**

We have systematically investigated the electronic properties calculations i.e., band structures and total density of states (DOS) of both the 2D monolayer pristine $MoS_2$ and Mn doped $MoS_2$ i.e. Mn-$MoS_2$. For the periodic 2D layer structure (i.e., 2D slab) computations, a single 2D TMDs (here both the $MoS_2$ and Mn-doped $MoS_2$) layer terminated on the ($10\bar{1}0$) (Mo-/Mn-edge) and ($\bar{1}010$) (S-edge) boundaries with three Mo per unit cell has been considered as shown in Figure 2. It should be mentioned here that the exposed surfaces are generally the (001) basal plane of the S−Mo−S (Mn-doped in the case of Mn-MoS$_2$) tri-layer, the Mo-/Mn-edge (10$\bar{1}$0) and S-edge ($\bar{1}$010). The rigid periodic structure computations and the equilibrium structures were obtained by performing hybrid dispersion corrected periodic density functional theory (in short DFT-D) i.e. here B3LYP-D3 method[50–59] implemented in *ab initio* based CRYSTAL17 suite code.[60] The electronic properties calculations were obtained by using the same B3LYP-D3 method.[61–64] We have performed spin polarized calculations to obtain the equilibrium structures and to study the electronic properties during periodic hybrid DFT-D calculations. A spin-polarized solution has been computed after definition of the (α, up spin and β, down spin) electron occupancy. In other words, it may be here noted that spin-unrestricted wave functions are used in the present calculations to incorporate spin polarization. This has been performed by using the keywords "ATOMSPIN" and "SPINLOCK" in *ab initio* CRYSTAL17 program.[60] In the present calculations, we have accounted for the weak long-range van der Waals (vdW) dispersion effects[65] resulting from the interaction between atoms by including the semi empirical corrections (Grimme's "–D3" corrections).[57] The weak vdW interaction between the layers of both the materials (MoS$_2$ and Mn-MoS$_2$) has been included in the present DFT calculations by adding Grimmes's semi-empirical dispersion parameters.[51–55] Triple-ζ valence with polarization function quality (TZVP) Gaussian basis sets were used for Sulphur (S)[66,67] and Manganese (Mn)[66] atoms, and HAYWSC-311 (d31) G type basis sets with Hay and Wadt small Effective Core Pseudopotentials (ECPs) for Molybdenum (Mo).[68] DFT-D method provides a good quality geometry of the 2D layered structure material after reducing the spin contamination effects such that it will not show any effect on the electronic structure and electronic properties calculations (i.e., band structure and the total density of states (DOS)).[38,54,69–72] The threshold used for evaluating the convergence of the energy, forces and electron density was set to 10$^{-7}$ a. u. for each parameter. The height of the unit cell was formally set to 500 Å (which considers there is no periodicity in the z-direction in the 2D slab model in CRYSTAL17 code), i.e., the vacuum region of approximately 500 Å was considered in the present calculations to accommodate the vacuum environment.[39,73] The unit cell of the 2D monolayer MoS$_2$ has been extended to a 3 × 3 × 1 to form a supercell and Mn atoms were doped by replacing the Mo atoms. It was found that the Mn-doping concentration was 12.5% in the 2D Mn-MoS$_2$ material (as the optimized supercell consisted of 9 Mo atoms out of which 1 Mo atom at the exposed edge was replaced with an Mn atom which leads the ratio Mn:Mo atoms to 1:8, thus making the doping concentration 12.5% near the desired active edges as shown in Figure 1). In the atomic structure relaxation simulation, a vacuum slab of 500 Å was inserted between the layers to avert the interlayer interaction.

The electronic band structures and total DOS calculations have been performed at the equilibrium structures of the TMDs by employing the same DFT-D method. All the integrations of the first Brillouin zone were sampled on 20×20×1 Monkhorst-pack,[74] k-mesh grids for the pristine 2D $MoS_2$ and 4×4×1 for 2D Mn-$MoS_2$. The k-vector path taken for plotting the band structure was selected as $\Gamma - M - K - \Gamma$ for both the materials (i.e., pristine $MoS_2$ and Mn-$MoS_2$). The atomic orbitals of Mo, S, and Mn were used to compute and plot the total DOS for the α electrons which is enough to describe the electronic properties of the 2D Mn-$MoS_2$ material. The single point calculation has been performed at the equilibrium geometry to form the normalized wave function at zero Kelvin temperature with respect to vacuum. To create the graphics and analysis of the crystal structures studied here, a visualization software VESTA[75] was used. We are aware that lateral interactions of the adsorbed species may change the free energies for different surface coverages. The same applies for different temperatures.[42,44,45]

## 2.2 Finite Cluster Modeling

Further, we developed a finite non-periodic molecular cluster model system for both the 2D monolayer pristine $MoS_2$ and Mn-$MoS_2$ materials to investigate HER mechanism by using GAUSSIAN 16[47] suite code. A non-periodic finite molecular cluster model $Mo_{10}S_{21}$ system for the pristine 2D monolayer $MoS_2$ and $Mn_1Mo_9S_{21}$ for the 2D monolayer Mn-$MoS_2$ TMD (as shown in Figure 1) has been considered here to investigate HER in both the gas phase and solvent phase calculations, and the M06-L[76,57] DFT method with a spin-unrestricted wavefunction has been applied to investigate the reaction pathways, kinetics, barriers, and mechanism. Figure 1 shows how we extract a triangular cluster from the periodic array to expose only the Mo edges. Schematic representation of the finite molecular cluster $Mn_1Mo_9S_{21}$ is shown in Figure 1. This M06-L DFT method is a technique used for energetics, equilibrium structures, thermochemistry, and frequency calculations of the molecular cluster structures, and it has been found that the M06-L method provides reliable energy barriers for reaction mechanisms of organometallic catalysts.[9,10,40,76–78] We focused on the energy barriers and changes of free energy during reaction to explore the reaction pathways by employing Minnesota density functional based on the meta-GGA approximation which is intended to be good and fast for transition metals.[76,77,79] We used 6-31G** Gaussian basis sets for H,[80,81] S,[82] O[83] and Mn[84] atoms, while LANL2DZ Gaussian basis set with the effective core potentials for Mo[85,86] atom. The transition state theory (TST) has been applied to located both the Volmer, Tafel and Heyrovsky transition states (TSs), and the OPT=QST2 and OPT=QST3 algorithms have been used to find out both the TSs which are implemented in GAUSSIAN 16[47] suite code. The transition structures or saddle points (Volmer, Tafel and Heyrovsky reaction steps) were computed to find the reaction barriers by confirming one imaginary frequency, modes of vibration, and intrinsic reaction coordinate (IRC) calculations.[9,10] Different transition states (TSs) were computed at optimized geometry and to visualize them, ChemCraft[87] was used. Moreover, the Heyrovsky reaction mechanism was studied by deliberately adding three water molecules and a hydronium ion in the vicinity of the intended reaction region. The water cluster model ($4H_2O + H^+$) was prepared as follows: 4 water molecules were placed adjacent to each other connected via hydrogen bond and a proton was attached to one of the water molecules. This model was prepared to simulate the reaction of $H_2$ formation during Heyrovsky process.

The two horizontal dashed lines indicate terminations along the ($10\bar{1}0$) Mn-/Mo-edge and ($\bar{1}010$) S-edge. The two triangles represent the terminations for Mn-/Mo-edge and S-edge clusters and the dangling bonds in the finite cluster have been set by considering a tringle as shown in Figure 2. Each Mo atom in the basal plane (001) of the finite molecular cluster model has oxidation state of +4 (and the oxidation state of the Mn atom is +4) and they are bonded with 6 S atoms (3 S at the upper plane and 3 S at the lower plane of Mo) which gives a contribution of 4/6 = 2/3 electrons towards each Mo-S bonding resulting a stabilized structure. The same can be understood with the oxidation state of S in the basal plane. Sulfur (S) atom has -2 oxidation state and bonding with 3 Mo atoms which results in a contribution of 2/3 electrons towards each Mo-S bond. Similarly, the edges of the periodic molecular cluster ($00\bar{1}0$) is being stabilized with the 2 local electron Mo-S bonds (as well as Mn-S bonds) having a single electron contribution towards each bond. So, at the edges each Mo atom contributes 2×1 electrons towards local Mo-S bonds plus 4×(2/3) electron contribution towards 4 Mo-S bonding in the basal plane as shown in the Figure 1. This 14/3 {i.e., (2×1) + [4×(2/3)]} contribution of electrons towards the Mo-S bonding of the edge Mo atom is satisfied with the *d²* configuration of one Mo atom and *d¹* configuration of two Mo atoms at the edges. This configuration leads the molecular system with the periodicity of 3 which results in the achievement of a stabilized molecular cluster model having three edges without any unsatisfied valency.[88]

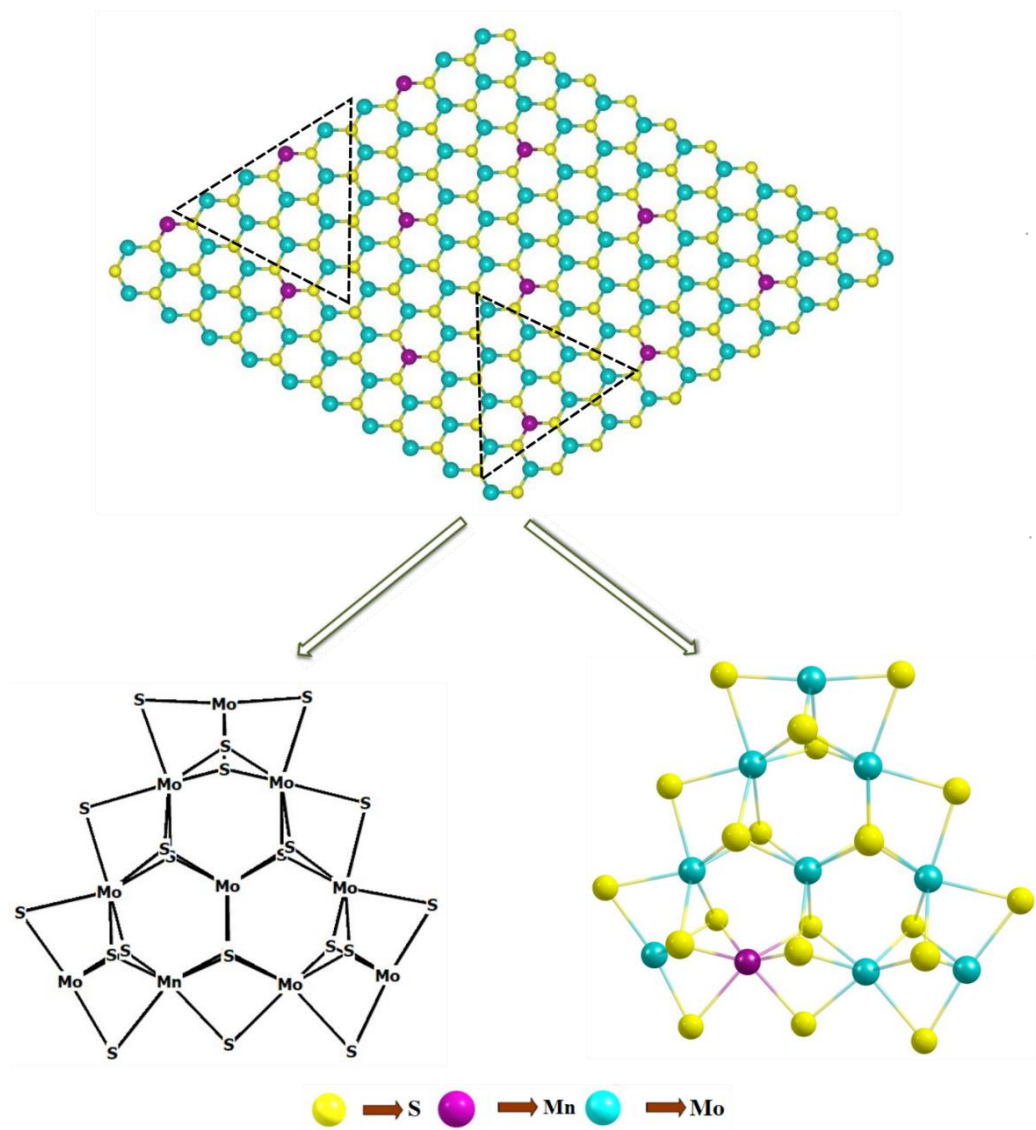

**Figure 1:** Non-periodic finite molecular cluster model ($Mn_1Mo_9S_{21}$) as derived from the 2D Mn-$MoS_2$ monolayer material (represented by the dotted triangle) is shown here.

To model the solvation effects, we used polarizable continuum model[89] (PCM) with water as a solvent. The PCM method using GAUSSIAN16[47] uses an external iteration method where the energy in the solution is computed by making the solvent reaction filed self-consistent with solute electrostatic potential, and as the real reaction is to take place in solvent phase. For specifying the molecular cavity used in PCM, we used the UAHF set i.e., United Atom Topological Model.[90] We modeled our reaction mechanism in water having static dielectric constant of 78.36 (zero-frequency, 298.15 K, 1 atm).[90,91] The geometric optimization and molecular energy in the solvent phase were calculated using the above-mentioned method.

## 2.3 Theoretical Calculations and equations

It is important here that we examine the equilibrium structures to find the free energy difference (here Gibbs free energy difference: ΔG i.e., relative Gibbs free energy) between the intermediate states, and ultimately the lowest-barrier pathway. All the HER steps have been explored with respect to standard hydrogen electrode (SHE). For each species, the free energy (G) can be expressed using the following equation:

$$G = E_{DFT} + E_{ZPE} + \int C_p \, dT - TS \qquad (1)$$

where $E_{DFT}$ is the ground state electronic energy, $E_{ZPE}$ is the zero-point vibrational energy, $C_p$ is the lattice specific heat capacity, $S$ is the entropy and T is the temperature (here 298.15K) which is kept constant throughout. The change in the free energy (ΔG), the change in enthalpy (ΔH) and the change in electronic energy (ΔE) of the reaction intermediates at pH = 0 has been calculated using the following equations:

$$\Delta G = \Sigma G_{Products} - \Sigma G_{Reactants} \qquad (2)$$

$$\Delta H = \Sigma H_{Products} - \Sigma H_{Reactants} \qquad (3)$$

$$\Delta E = \Sigma E_{Products} - \Sigma E_{Reactants} \qquad (4)$$

For all purposes, we have considered the standard hydrogen electrode (SHE) condition where electrons (e⁻) and protons (H⁺) (pH = 0) are in equilibrium with 1 atm $H_2$. The change of Gibbs free energy of an electron at SHE condition was computed by considering the difference between the free energies of half of the $H_2$ molecule and a proton (H⁺) when pH=0. Tafel slope (***b***) gives information about the kinetics, rate determining steps of the electrochemical reaction, the energy required to achieve activity, etc. The Tafel slope (***b***) has been computed by using the formula ***b*** = 2.303 RT/nF; where F is Faraday constant, and n is number of electrons involved in the subject reaction.[88] Tafel slope is an inverse measure of how strongly the reaction rate responds to changes in potential.

# 3 Results and Discussions

## 3.1 Periodic vacuum slab inference

The equilibrium lattice parameters and average bond distances are listed in Table 1. The lattice constants (*a* and *b*) and the average Mo-S bond distance of the pristine 2D monolayer $MoS_2$ obtained by the DFT-D method are consistent with the previous reported results.[69] The values of lattice constants (*a = b)* are 3.18 Å and the Mo-S bond distance is about 2.41 Å which accord with the previous reported values, and it has a hexagonal 2D layer ***P-6m₂*** symmetry.[92] This is a good estimation as compared to a work where the 2D monolayer $MoS_2$ structure was doped with 4% impurity and the bond distance was 2.39 Å between the Mn and the nearest S atom.[92] The doping of transition metal in the 3×3×1 supercell of the 2D monolayer $MoS_2$ has changed its symmetry from ***P-6m₂*** to ***P1*** when the 2D Mn-$MoS_2$ has been formed. The average bond distance between the Mn and the nearest S atom was computed to be 2.30 Å which agrees well the previous result within 0.09 Å.[69] From the electronic properties calculations obtained by the same DFT-D method, we observed a direct band gap of 2.6 eV at *K* point in the Brillouin zone of the pristine 2D single layer $MoS_2$ material as shown in Figure 2a which is well harmonized with the previous theoretical and experimental results. The computed electronic band gap is slightly lower than the band gap obtained by the GW approximation of the 2D monolayer $MoS_2$ TMD which was 2.8 eV.[93] The Fermi level (**E_F**) was found at -6.36 eV depicted in the non-normalized band structure and DOS calculation as shown in Figure 2a highlighted by dotted blue color. After Mn-doping in the pristine 2D monolayer $MoS_2$ material, the band structures have been changed i.e., the Fermi level (**E_F**) shifted to -5.04 eV (with respect to the **E_F** of the 2D pristine $MoS_2$) and it was computationally found that the bands are overlapped around the **E_F** as shown in Figure 2b. The present DFT-D study shows that the Fermi level was found at -5.04 eV in the case of the 2D monolayer Mn-$MoS_2$ with zero band gap indicating the conducting character of the material. In other words, this zero-band gap suggests that the Mn-doping in the pristine TMD makes the 2D semi-conducting $MoS_2$ material into a conducting material in nature. This can also be justified by computing the electron density contribution from the 3*d*-subshells of the Mn atoms doped in the 2D monolayer $MoS_2$ material (as it can be seen from the *d* subshell DOS at the right-hand side in Figure 2b). In other words, due to the addition of Mn atoms in the pristine $MoS_2$ to form the 2D monolayer Mn-$MoS_2$ material, the electronic band gap of the Mn-$MoS_2$ was decreased to zero depicted in the band structures and DOS calculations in Figure 2b. The addition of Mn to the pristine 2D TMD $MoS_2$ changes the electron accumulation in the bands as shown in the DOS calculations suggesting high electron mobility with an indication of possibly good catalytic activities for HER.

**Table 1:** Lattice Parameters of both the 2D monolayer pristine MoS$_2$ and Mn-MoS$_2$ TMD materials computed by the hybrid periodic DFT-D method have been provided here.

| System | Lattice constants (a=b) | Interfacial angles (α, β and γ) | Space group symmetry | Average bond distance | |
|---|---|---|---|---|---|
| | | | | Mo-S | Mn-S |
| MoS$_2$ | 3.180 Å | α = β = 90.0° and γ = 120.0° | *P-6m$_2$* | 2.411 Å | --------- |
| Mn-MoS$_2$ (3x3 supercell) | 9.451 Å | α = β = 90.0° and γ = 120.0° | *P1* | 2.409 Å | 2.303 Å |

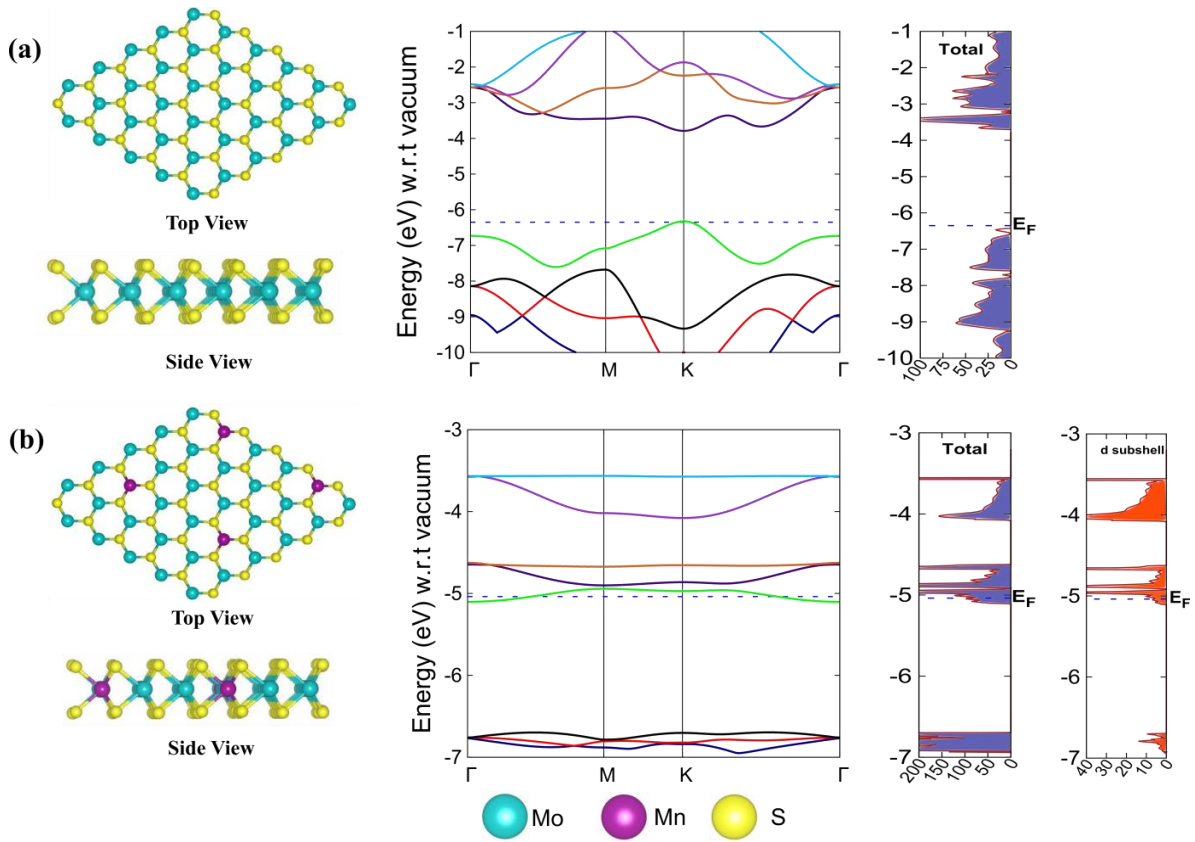

**Figure 2: (a)** Top view and side view of the pristine 2D monolayer pristine MoS$_2$ with its band structure and total DOS; **(b)** Top view and side view of the 2D monolayer Mn-MoS$_2$ with band structure and total DOS along with the contributing component of the *d*-subshell DOS of the Mn atom in the total DOS are displayed here.

Now it is confirmed from the electronic structure and properties calculations that the 2D monolayer Mn-$MoS_2$ TMD material has shown conducting properties (i.e., conducing in nature), so we proceed with our study in the direction of theoretical and computational development for optimum electrocatalyst. Hydrogen production from electrolysis of water is conceived as a match for the growing need to alternate green energy sources. In this regard, we carefully audit the fundamentals of HER and control parameters of the kinetics of reaction. We have outlined molecular simulation approaches that will help us to tackle the challenges that lie at hand in design of cheap and practical catalysts.

## 3.2 HER recapitulation

Now, we turned our attention to investigate the detailed HER mechanism by predicting energetics for the various reaction steps relevant to the HER in the case of 2D monolayer Mn-$MoS_2$ material. Using the molecular cluster model system of the 2D monolayer Mn-$MoS_2$ material, we can now add or subtract electrons ($e^-$) and protons ($H^+$) independently in discrete $H_2$ evolution reaction steps. First, we calculate the free energies of the most likely intermediates to serve as a basis for describing the thermodynamics of HER. Then, we examine the barriers of the various reaction steps to locate the rate limiting step during the reactions. In general, HER is a two-way reaction mechanism, and the most generally accepted reaction mechanism is given as follows: HER can occur via the ***Volmer-Heyrovsky*** process or the ***Volmer-Tafel*** process as depicted in Figure 3. The process begins with water adsorption and dissociation where at first, the $H_2O$ reacts with electron ($e^-$) to produce $H^+$ and $OH^-$ which takes place at the active site of the electrocatalyst (more specifically the active surface of the electrocatalyst). The further mechanism can occur either through Heyrovsky reaction step or the Tafel step. In the Heyrovsky reaction, the adsorbed hydride ion reacts with the adjacent water molecule or more specifically with the solvated proton of the adjacent water to produce $H_2$. In the Tafel reaction step, where two adsorbed hydrogens are adjacent to each other, recombine to form $H_2$ during the reaction.

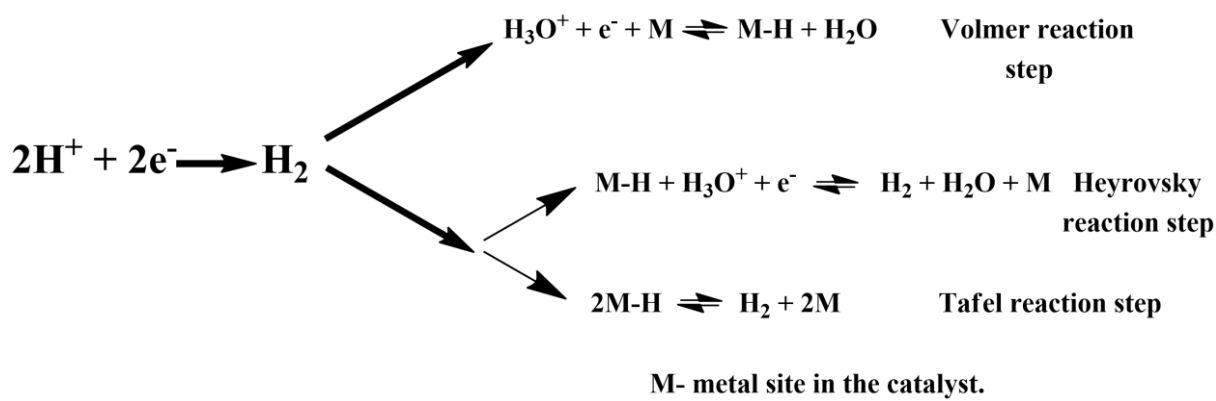

**Figure 3:** Possible reaction pathways of HER in acidic electrolyte is presented here.

Generally, it has been found that the Volmer-Heyrovsky reaction mechanism is more likely to be predominant when transition metal-based catalysts are used because of their good adsorption free energy.[94] So, the analysis of the proposed catalyst for its involvement in the mechanics and kinetics of the reaction (as mentioned in Figure 3) was put into the effects. To study the HER mechanism, we computationally developed a cluster model system for the 2D monolayer Mn-MoS$_2$ and performed non-periodic M06-L DFT theory. The two processes have been carefully studied and discussed further.

## 3.3 Volmer-Heyrovsky Mechanism

The Volmer-Heyrovsky reaction pathway in the vicinity of the active site of the 2D monolayer Mn-MoS$_2$ TMD has been schematically presented in Figure 4. This process is a multistep electrode reaction which has been described and the reaction steps above, intermediates and transition states (TSs) occurred during the HER process have been reported here.

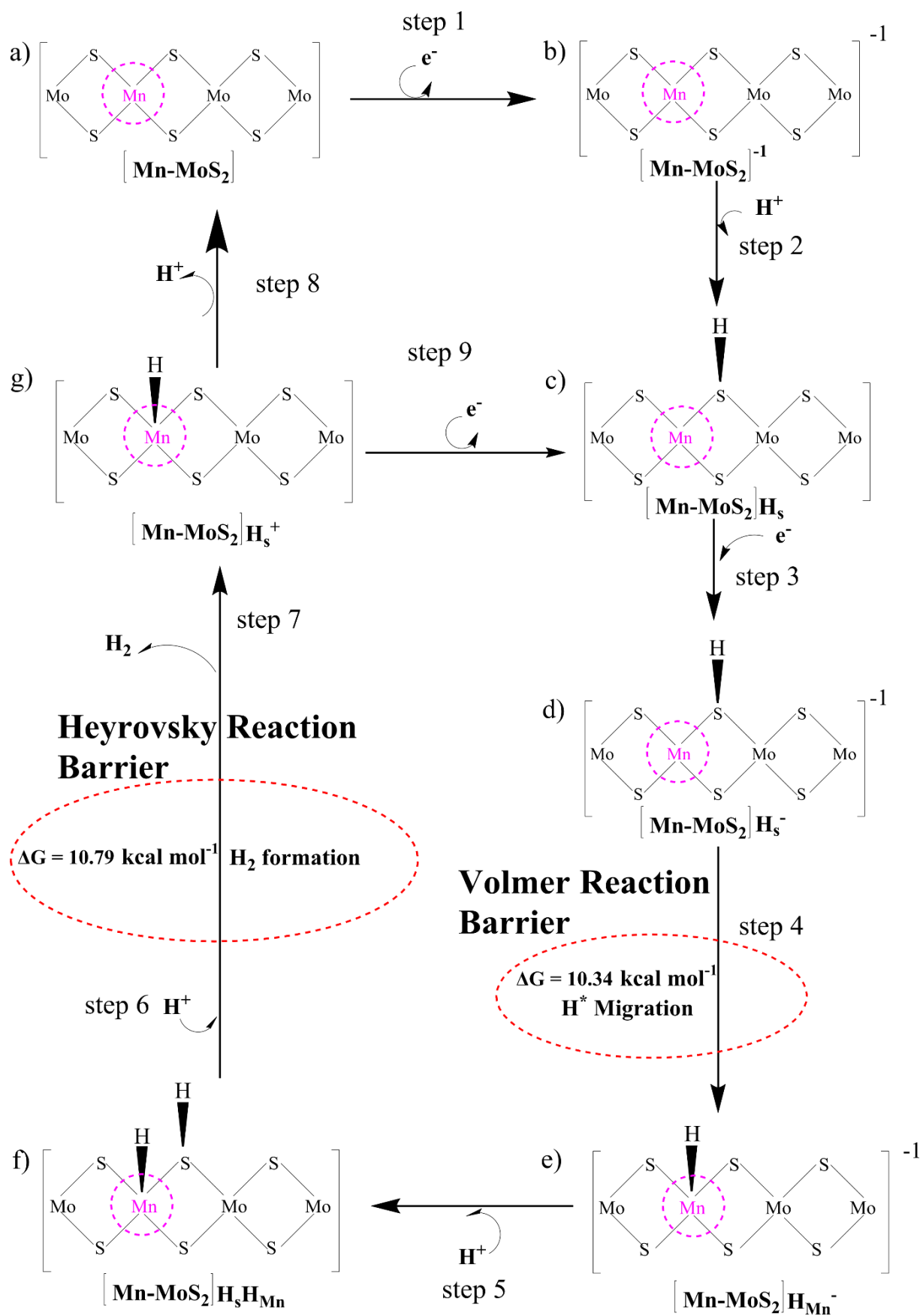

**Figure 4:** Proposed reaction pathway scheme for HER using 2D monolayer Mn-MoS$_2$

electrocatalyst is shown here.

In order to evolve an $H_2$ molecule, protons and electrons must be added to the 2D Mn-$MoS_2$ molecular cluster $Mn_1Mo_9S_{21}$. Here, it is useful to examine first the most stable structures with each number of extra electrons and each number of extra protons to understand the free energy differences between intermediate states, and ultimately find the lowest-barrier pathway. A detailed description of the HER process involved in the subject reaction is required to explain the electrochemistry. The first process is the dissociation of water in Heyrovsky's reaction step. To initiate the HER, one electron is absorbed on the surface of the 2D monolayer Mn-$MoS_2$ TMD as depicted in Figure 4. This step takes place at the SHE conditions which is the footing of the thermodynamic potentials for oxidation and reduction processes. The first reduction is achieved with a reduction potential ΔG about -129.66 mV resulting into [Mn-$MoS_2$]$^-$ from [Mn-$MoS_2$] with the changes of enthalpy (ΔH) and electronic energy (ΔE) about -2.93 and -2.97 kcal mol$^{-1}$, respectively, as shown in Table 2. Thereafter, $H^+$ from the solvent medium is adsorbed on the sulfur site which is the most energetically conducive site for the moment, forming [Mn-$MoS_2$]$H_s$ solvated cluster (as the first adsorption of H at the Mn site with an energy cost ΔG = 3.57 kcal mol$^{-1}$ so the lower barrier path is to follow the [Mn-$MoS_2$]$^-$ → [Mn-$MoS_2$]$H_s$ rather than [Mn-$MoS_2$]$^-$ → [Mn-$MoS_2$]$H_{Mn}$ path). In the follow up step, [Mn-$MoS_2$]$H_s^{-1}$ complex is formed due to the addition of another electron from the solvent with a second reduction potential of -199.91 mV. Hence, we are reporting this HER as a two-electron transfer reaction. In the next step, the hydride ($H^-$) ion from the sulfur site migrates over to the neighboring responsive Mn site. The migration of H$^•$ from the S to Mn site is the H$^•$-migration Volmer reaction step. This transition structure i.e., the Volmer transition state (TS) or H$^•$-migration reaction TS (TS1) is corroborated by the detection of imaginary vibrational frequency at the site of transition of hydride ion from S to Mn. The formation of TS is accompanied by positive free energy change of 7.23 kcal mol$^{-1}$ in the gas phase calculations. $H^+$ from medium again attacks; from here either the Tafel or the Heyrovsky process can take place. The Heyrovsky part is further shown in Figure 4. Computationally, we explicitly added 4$H_2O$-$H^+$ cluster (i.e., more specifically 3 water molecules and one hydronium ion ($H_3O^+$)) in vicinity of the active site of the catalyst and observed according to the simulation that the $H^-$ from the metal site (here Mn) and the $H^+$ from the water cluster (i.e., adjacent hydronium ion) creates a bond and then it evolves as $H_2$. This process is called Heyrovsky process, often mentioned as the Heyrovsky transition state (TS2) in reaction mechanism during HER. The Mn-S bond distance during TS2 was recorded to be 2.311 Å which is higher than the strain free bond distance of 2.303 Å as mentioned in Table 1. Finally, $H^+$ ($H_3O^+$) splits off to form the neutral Mn-MoS$_2$ surface. Alternatively, an electron is absorbed to obtain neutral [Mn-MoS$_2$]H$_s$. All the optimized reaction intermediates, reactants and products with TSs of our proposed reaction scheme have been illustrated in Figure 5.

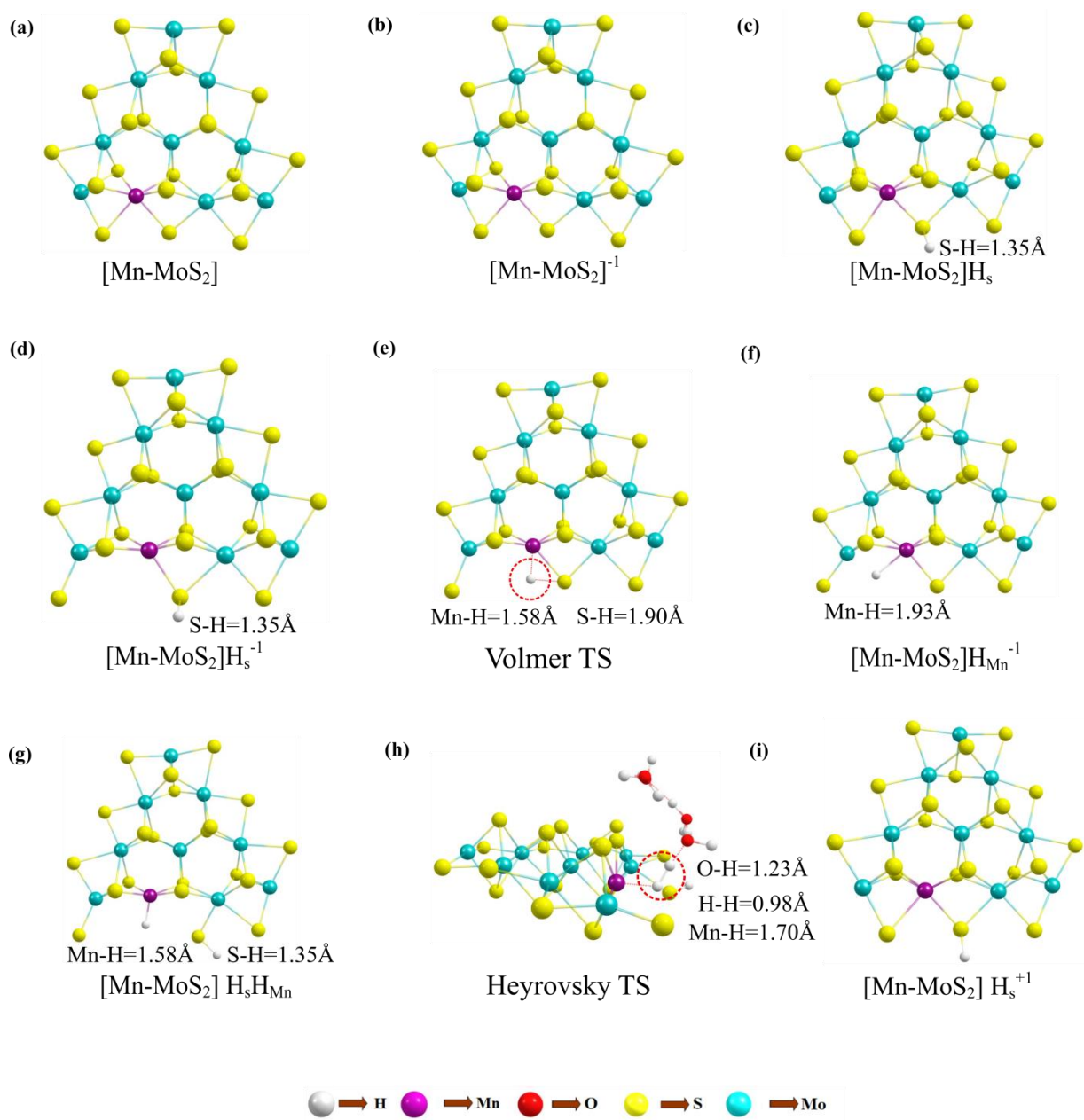

**Figure 5:** Equilibrium geometries of important reaction intermediates and TSs: (a) [Mn-MoS$_2$], (b) [Mn-MoS$_2$]$^{-1}$, (c) [Mn-MoS$_2$]H$_s$, (d) [Mn-MoS$_2$]H$_s^{-1}$, (e) Volmer TS, (f) [Mn-MoS$_2$]H$_{Mn}^{-1}$, (g) [Mn-MoS$_2$]H$_s$H$_{Mn}$, (h) Heyrovsky TS and (i) [Mn-MoS$_2$] H$_s^{+1}$ computed by M06-L DFT method considering the molecular cluster model system Mn$_1$Mo$_9$S$_{21}$ to represent 2D monolayer

of Mn-MoS$_2$ are shown here.

We keenly followed the different reaction pathway schemes as shown in Figure 4 but kept our focus on the two important saddle points i.e., Volmer reaction where an H$^\bullet$ atom migrates from the S to the transition metal site (here Mn site) i.e. H$^\bullet$-migration reaction TS1 and the other being Heyrovsky reaction step where H$^+$ from the adjacent water cluster and the H$^-$ from the Mn site recombine to form H$_2$. It has been observed that the Heyrovsky reaction step is the rate determining steps of HER for our system of interest (i.e., 2D monolayer Mn-MoS$_2$). The changes of free energy ($\Delta$G), enthalpy ($\Delta$H) and electronic energy ($\Delta$E) at each reaction step during HER process in gas phase calculations are listed in Table 2. In summary, we observed that the Volmer activation barrier is about 7.23 kcal mol$^{-1}$ and the H$_2$ formation reaction barrier i.e., Heyrovsky's reaction barrier is about 10.59 kcal mol$^{-1}$ computed in gas phase. The values of the activation barriers for the respective steps in gas as well as in solvent phases i.e., the values of $\Delta$G during the TSs formation are summarized and listed in Table 3. The changes of enthalpy ($\Delta$H) and electronic energy ($\Delta$E) during the formation of the TS1 i.e., H$^\bullet$-migration TS are 8.03 kcal mol$^{-1}$ and 8.17 kcal mol$^{-1}$, and similarly, the values of $\Delta$H and $\Delta$E during the H$_2$-formation i.e., TS2 are 10.63 kcal mol$^{-1}$ and 10.69 kcal mol$^{-1}$, respectively, reported in Table 2.

**Table 2:** The changes of free energy ($\Delta$G), enthalpy ($\Delta$H) and electronic energy ($\Delta$E) during H$_2$ evolution reaction mechanism in the gas phase at T = 298.15K and 1atm pressure are reported here. The units are expressed in kcal mol$^{-1}$.

|  | Reaction Intermediates | $\Delta$E (kcal mol$^{-1}$) | $\Delta$H (kcal mol$^{-1}$) | $\Delta$G (kcal mol$^{-1}$) |
|---|---|---|---|---|
| Step 1 | [Mn-MoS$_2$] $\longrightarrow$ [Mn-MoS$_2$]$^{-1}$ | -2.93 | -2.97 | -2.99 |
| Step 2 | [Mn-MoS$_2$]$^{-1}$ $\longrightarrow$ [Mn-MoS$_2$]H$_s$ | -26.98 | -27.21 | -27.25 |
| Step 3 | [Mn-MoS$_2$]H$_s$ $\longrightarrow$ [Mn-MoS$_2$]H$_s^{-1}$ | -2.48 | -2.91 | -4.61 |
| TS1 |  | 8.17 | 8.03 | 7.23 |
| | [Mn-MoS$_2$]H$_s^{-1}$ ⟶ Volmer TS | | | |
| Step 4 | Volmer TS ⟶ [Mn-MoS$_2$]H$_{Mn}^{-1}$ | -36.36 | -36.41 | -36.49 |
| Step 5 | [Mn-MoS$_2$]H$_{Mn}^{-1}$ ⟶ [Mn-MoS$_2$]H$_s$H$_{Mn}$ | -34.67 | -35.00 | -35.19 |
| Step 6 | [Mn-MoS$_2$]H$_s$H$_{Mn}$ ⟶ [Mn-MoS$_2$]H$_s$H$_{Mn}$ + 4H$_2$O_H$^+$ | 0.04 | 0.02 | 0.01 |
| TS2 | [Mn-MoS$_2$]H$_s$H$_{Mn}$4H$_2$O_H$^+$ ⟶ Heyrovsky TS | 10.69 | 10.63 | 10.59 |
| Step 7 | Heyrovsky TS ⟶ [Mn-MoS$_2$] H$_s^{+1}$ | -28.20 | -28.83 | -29.38 |

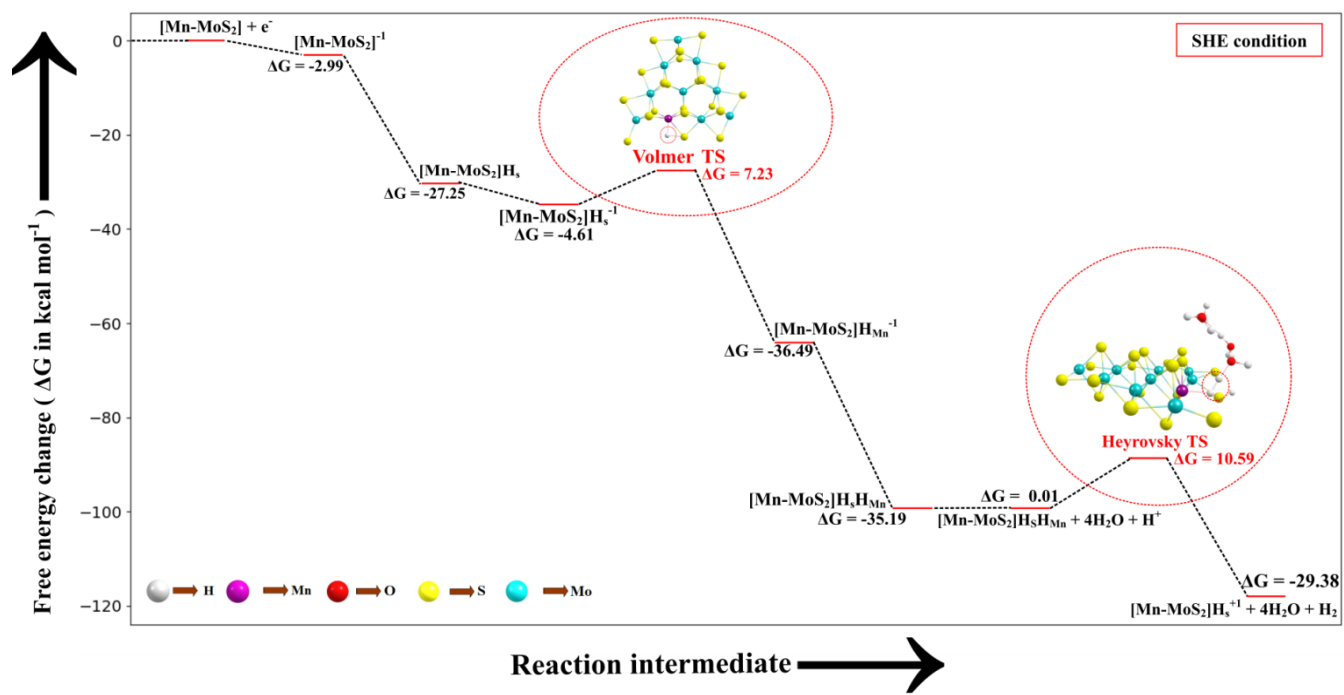

**Figure 6:** Free energy diagram i.e. potential energy surfaces (PES) of the Volmer-Heyrovsky reaction pathway on the surface of the 2D monolayer Mn-MoS$_2$ material computed in the gas phase is presented here.

**Table 3:** Activation reaction energy barriers i.e., the changes of free energy (ΔG), enthalpy (ΔH), and electronic energy (ΔE) in HER process on the surfaces of the 2D monolayer Mn-

MoS$_2$ TMD are reported here. The units are expressed in kcal mol$^{-1}$.

| Activation Barrier | ΔG (kcal mol$^{-1}$) in gas phase | ΔE (kcal mol$^{-1}$) in solvent phase | ΔH (kcal mol$^{-1}$) in solvent phase | ΔG (kcal mol$^{-1}$) in solvent phase |
|---|---|---|---|---|
| **Volmer reaction barrier** | 7.23 | 11.84 | 9.60 | 10.34 |
| **Heyrovsky reaction barrier** | 10.59 | 12.56 | 10.36 | 10.79 |

PCM calculations have been formed to incorporate the solvation effects during the HER. From our present DFT calculations considering the PCM system, we have predicted that the Volmer step energy barrier i.e., the reaction barrier (ΔG) of the TS1 is about 10.34 kcal mol$^{-1}$ in the solvent phase with the change of electronic energy (ΔE) about 11.84 kcal mol$^{-1}$ reported in Table 3. Recently Yu et al.[9] reported that the TS1 H$^{\bullet}$-migration reaction energy barriers during Volmer step in the case of the 2D monolayer pristine MoS$_2$, WS$_2$ and the hybrid W$_{0.4}$Mo$_{0.6}$S$_2$ TMD alloy are about 17.7 kcal mol$^{-1}$, 18.1 kcal mol$^{-1}$ and 11.9 kcal mol$^{-1}$, respectively, computed in the solvent phase by considering PCM. Moreover, the Heyrovsky TS barriers for the pristine MoS$_2$, WS$_2$ and W$_{0.4}$Mo$_{0.6}$S$_2$ TMD alloy were calculated as 23.8 kcal mol$^{-1}$, 21.3 kcal mol$^{-1}$ and 13.3 kcal mol$^{-1}$, respectively. For the case of the 2D monolayer Mn-MoS$_2$ material, the calculated value of ΔG during Heyrovsky's reaction step TS2 was 10.79 kcal mol$^{-1}$ with the change of electronic energy 12.56 kcal mol$^{-1}$ in the solvent phase calculation. The changes of enthalpies (ΔH) during the formation of the TSs TS1 and TS2 computed in the solvent phase are about 9.60 and 10.36 kcal mol$^{-1}$, respectively, obtained by the M06-L DFT method. The energy barriers for different materials are mentioned in Table 4. We reported from our DFT calculations that the proposed catalyst shows much lower activation energies during HER, so the 2D monolayer Mn-MoS$_2$ can be characterized as a highly efficient HER catalyst. In other words, the present DFT calculations reveal that both the H$^{\bullet}$-migration (TS1) and Heyrovsky's reaction (TS2) barriers during HER process on the active surfaces of 2D monolayer Mn-MoS$_2$ TMD material are the lowest compared to the other TMDs and their hybrid alloys indicating an excellent electrocatalyst for effective H$_2$ evolution.

**Table 4:** Comparison of Reaction Energy Barriers (ΔG) in HER Solvent phase using different catalysts are reported here. The units are expressed in kcal mol$^{-1}$.

| Material | Volmer Reaction Barrier (ΔG) (kcal mol$^{-1}$) | Heyrovsky Reaction Barrier (ΔG) (kcal mol$^{-1}$) | References |
|---|---|---|---|
| $MoS_2$ | 17.7 | 23.8 | 9 |
| $WS_2$ | 18.1 | 21.3 | 9 |
| $W_{0.4}Mo_{0.6}S_2$ | 11.9 | 13.3 | 9 |
| $Mn-MoS_2$ | 10.30 | 10.8 | This work |

The comparison of different electrocatalysts based on the activation barriers in the reaction computed in the solvent phase can be visualized by the graphical illustration below shown in Figure 7. These figures depict that the 2D monolayer Mn-MoS$_2$ TMD material shows the least activation barriers (i.e., both the Volmer and Heyrovsky reaction barriers) among the others operational electrocatalysts mentioned in Table 4.

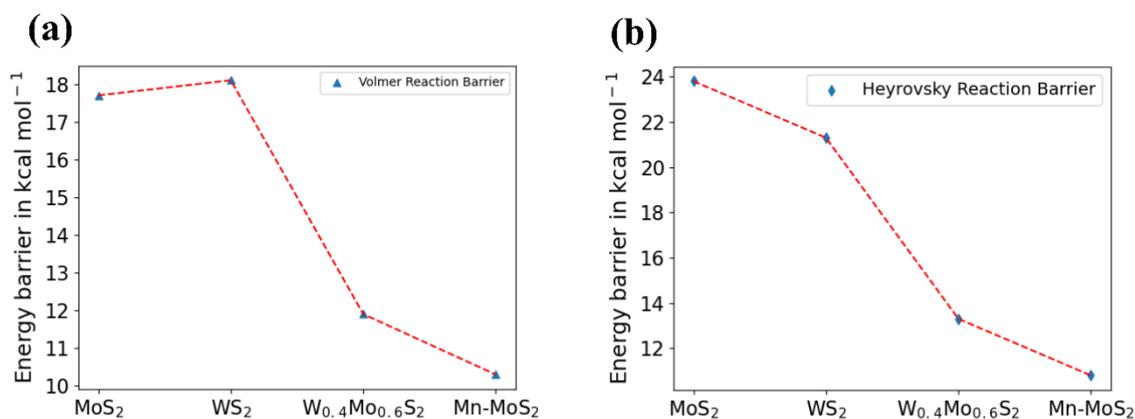

**Figure 7:** Graphical illustration of (a) Volmer and (b) Heyrovsky reaction barrier in solvent phase for different materials is depicted here.

From the transition state theory[95] (TST) or the activated complex theory by including the DFT calculations, we have determined the turnover frequency (TOF) for H$_2$ evolution per edge of Mn atom in the 2D monolayer Mn-MoS$_2$ catalyst. For the theoretical determination, we used the formula: rate = ($K_B$T/h) × exp(-ΔG/RT);[88] where $K_B$ is the Boltzmann constant, T (here 298.15 K) is temperature, h is the Planck's constant, R is the universal gas constant and ΔG corresponds to the free energy barrier. The TOF obtained for the 2D monolayer Mn-$MoS_2$ from the $H_2$ formation reaction energy barrier in the Heyrovsky mechanism (in solvent phase) is about 7.74 x $10^4$ $sec^{-1}$. The high value of TOF is suitable for an efficient $H_2$ evolution during the reaction. For example, the excellent and well-functioning electrocatalyst developed by Yu et.al., (hybrid $W_{0.4}M_{0.6}S_2$ alloy material) showed TOF value as high as 1.1×$10^3$ $sec^{-1}$. The TOF values of other practical catalysts such has the 2D monolayer $MoS_2$, $WS_2$, etc., have been mentioned in Table 5 for comparison. 2D Mn-$MoS_2$ material showed quite a high TOF value which aids the fact that this material will show excellent and efficient performance during HER.

**Table 5:** Heyrovsky's reaction barrier (ΔG) and Turnover frequency (TOF) for the 2D monolayer $MoS_2$, $WS_2$, $W_{0.4}Mo_{0.6}S_2$ and Mn-$MoS_2$ TMDs in solvent phase are tabulated here.

| Material | Barrier in gas phase (ΔG) (kcal $mol^{-1}$) | Barrier in solvent phase (ΔG) (kcal $mol^{-1}$) | Turnover frequency (TOF) in solvent phase ($sec^{-1}$) | Reference |
|---|---|---|---|---|
| $MoS_2$ | 16.0 | 23.8 | 2.1 × $10^{-5}$ | 9 |
| $WS_2$ | 14.5 | 21.3 | 1.5 × $10^{-3}$ | 9 |
| $W_{0.4}Mo_{0.6}S_2$ | 11.5 | 13.3 | 1.1 × $10^3$ | 9 |
| Mn-$MoS_2$ | 10.6 | 10.8 | 7.74 × $10^4$ | This work |

    We can observe that the 2D monolayer Mn-$MoS_2$ shows comparable results to the 2D monolayer hybrid $W_{0.4}Mo_{0.6}S_2$ alloy material. Therefore, it can be said that the 2D Mn-$MoS_2$ can prove to be a better and practical alternative for superb catalytic performance for HER. Another electrochemical parameter i.e., Tafel slope (which gives information about the kinetics, rate determining steps of the electrochemical reaction and the energy required to achieve activity) is also one of the important factors to assess the performance of electro-catalysts. The experimentally observed Tafel slope (***b***) for the 1T phase of $MoS_2$ and $WS_2$ is about 40 mV $dec^{-1}$ and 55 mV $dec^{-1}$, respectively (synthesized via lithium intercalation at room

temperature).[96,97] The Tafel slope of other efficient and functional catalyst such as $MoS_2$ nanoparticles grown on graphene is 41 mV dec$^{-1}$ where electrochemical desorption was the rate limiting step during hydrogen catalysis.[22] The Tafel slope for hybrid $W_{0.4}Mo_{0.6}S_2$ alloy material synthesized via wet chemical route has been reported as 38.7 mV dec$^{-1}$.[9] The Tafel slope can also be calculated theoretically by taking into consideration the number of electrons transferred during HER. As stated earlier that the proposed reaction is a two-electron transfer mechanism, so our DFT-D computed Tafel slope turned out to be 29.58 mV dec$^{-1}$ for the 2D Mn-$MoS_2$ which is 9.12 mV dec$^{-1}$ lower than that of hybrid $W_{0.4}Mo_{0.6}S_2$ alloy material indicating that 2D Mn-$MoS_2$ is an excellent electrocatalyst for HER.

Our present computations are in strong favor of low energy barriers of both the Volmer and Heyrovsky steps in HER using the 2D monolayer Mn-$MoS_2$ which promises a favorable candidate for HER electrocatalyst. To further support our development, we implemented Natural Bond Orbital (NBO), highest occupied molecular orbital (HOMO) and lowest unoccupied molecular orbital (LUMO) calculations in DFT analysis. These calculations were performed in order to show appropriate perspective of $H_2$ formation at the active site from electronic charge and molecular orbital overlapping point of view. Precise Lewis structures i.e., structures which have maximum electronic charge in the Lewis orbitals, can be found out by calculating NBO. This study conveys interaction density or the overlap density from the wavefunctions. The solution to the multielectron atomic system requires an approximation called the linear combination of atomic orbitals (LCAO approximation). The qualitative picture of molecular orbital is analyzed by expanding the molecular orbital into any complete basis set of all atomic orbitals of nuclei. So, the multi-electron wavefunction in a molecule at a specific configuration of the nuclei can be given by expanding the orbital approximation to molecules. The wave function obtained from the NBO calculations is a linear combination of the atomic orbitals of the Mn, S, Mo and H atoms for the Volmer TS and the Mn, S, Mo, H and O atoms for Heyrovsky TS. The HOMO LUMO was obtained from the optimized transition structures (both Volmer and Heyrovsky transition states TS1 and TS2) as shown in Figure 8. The red color represents in phase bonding of the orbitals and the blue color shows out of phase bonding. The boundary value outlining the isosurface shown in Figure 8 was set at 0.009. The interval of values in which the isosurface is colored (from blue to red) was set from -0.1 to 0.1.

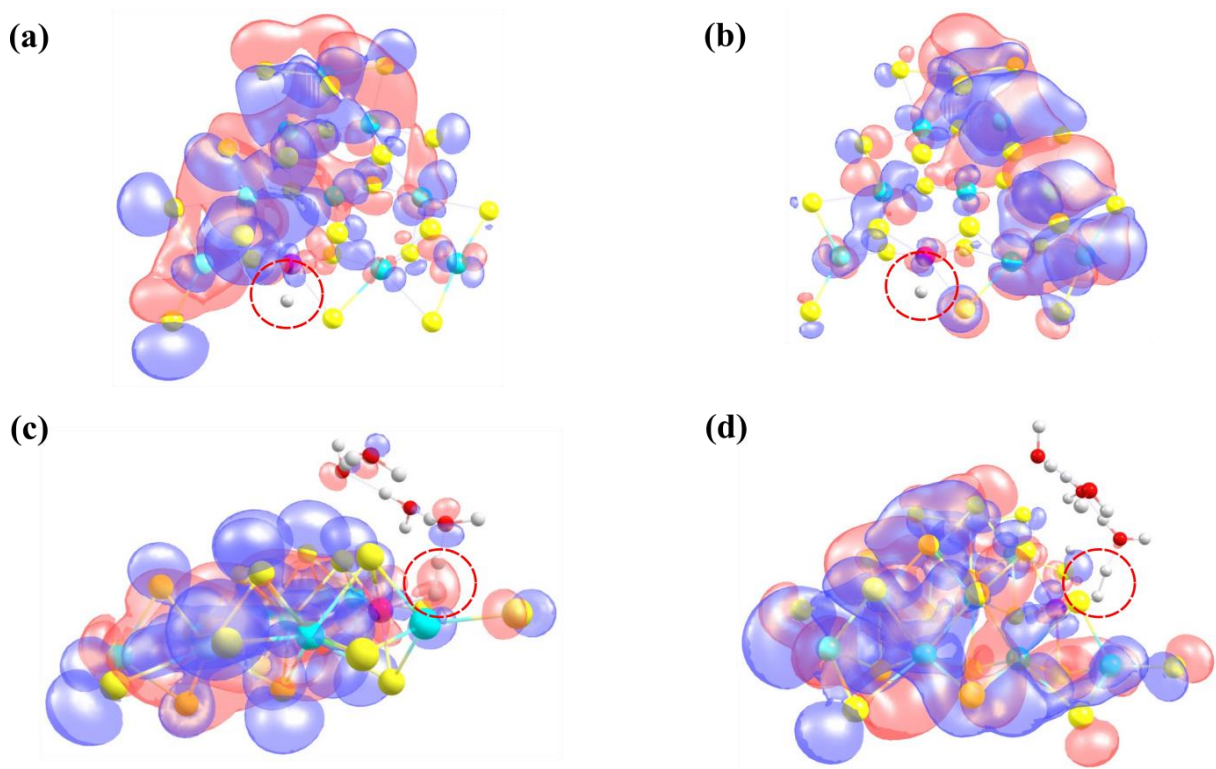

**Figure 8:** (a) HOMO of the Volmer TS (b) LUMO of the Volmer TS (c) HOMO of the Heyrovsky TS (d) LUMO of the Heyrovsky TS are shown here. The molecular orbitals involved in the HER and the position of hydrogen are highlighted by red dotted circle.

The insight can be drawn on the role of electronic structure in HER mechanism from the HOMO calculations of the transition states and the $H_2$ formation in a steady state due to better overlap of the *d*-orbital of Mn atom and the *s*-orbital of $H_2$ compared to the pristine 2D monolayer $MoS_2$ or $WS_2$. Therefore, one conclusion can be drawn here that in the rate limiting step of HER i.e., the Heyrovsky step, the stabilization of the atomic orbitals is also one of the key features for reducing this reaction barrier. The electron cloud around the H atoms in Heyrovsky TS is highlighted by red dotted circle (Figure 8 (c)). This step is backed up by the overlap of the atomic orbitals of H and Mn atoms along with $H_3O^+$ ion when $H_2$ is evolved. This is also one of the reasons why the 2D monolayer Mn-$MoS_2$ shows excellent activity for HER. The energy difference between HOMO and LUMO, also known as HOMO-LUMO gap is used to predict the stability transition metal based complex[98, 99] as it is the lowest energy electronic excitation that is possible in the molecule.

## 3.4 Volmer-Tafel mechanism

The proposed Volmer-Tafel reaction scheme is illustrated in Figure 9.

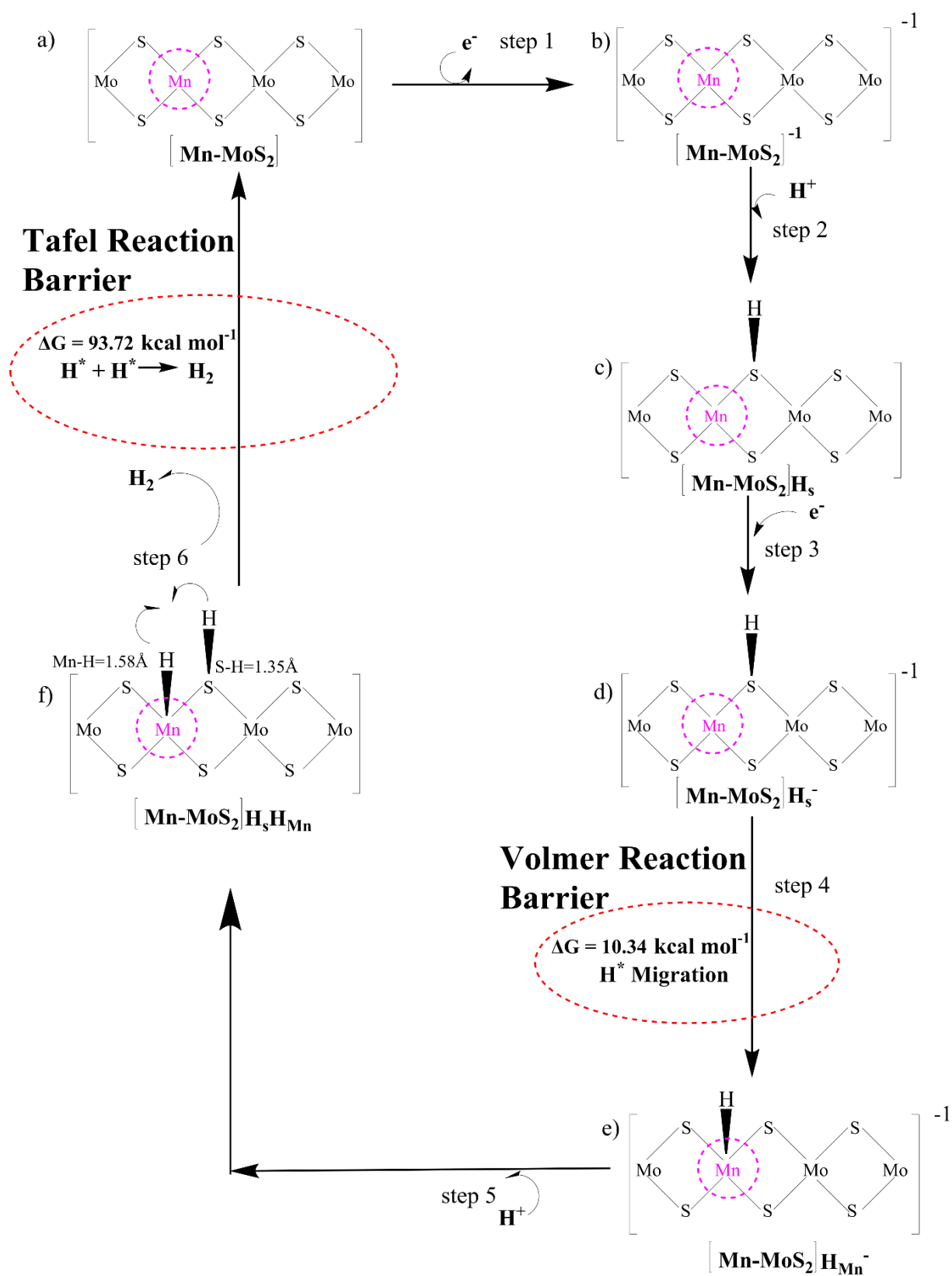

**Figure 9:** Reaction scheme with the possible pathway for Volmer – Tafel mechanism is shown here.

The Volmer-Tafel reaction steps are similar till the formation of [Mn-MoS$_2$]H$_s$H$_{Mn}$ complex. From here the process takes place as follows; two adsorbed hydrogens on the surface of the catalyst combine to evolve as H$_2$. The equilibrium structure of the Volmer – Tafel transition state (TS3) can be seen in Figure 10 a.

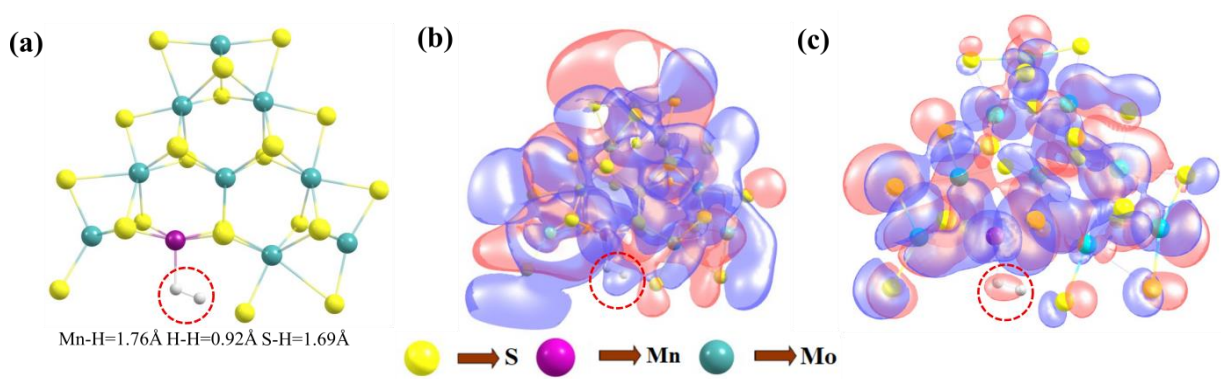

**Figure 10:** (a) Equilibrium structure of the Tafel TS, (b) HOMO and (c) LUMO of Tafel TS are shown here.

In this reaction, the adsorbed hydrogen at the sulfur site and the adsorbed hydrogen at the transition metal site combine to form H$_2$ (2H$^*$ → H$_2$; where H$^*$ represents hydrogen adsorbed on the active site of the catalyst). The Tafel reaction barrier (ΔG) in gas phase was recorded to be 90.13 kcal mol$^{-1}$ and 93.72 kcal mol$^{-1}$ in the solvent phase which are extremely higher than Volmer-Heyrovsky reaction barrier. The changes of enthalpy (ΔH) and electronic energy (ΔE) during H$_2$ formation in Tafel reaction step are 92.27 kcal mol$^{-1}$ and 93.66 kcal mol$^{-1}$, respectively, computed by the DFT method reported in Table 6. The Tafel reaction barrier (ΔG) of the TS3 is higher than the Heyrovsky's reaction barrier of the TS2 indicating that the Volmer-Tafel reaction step is thermodynamically less favorable. The HOMO-LUMO calculations were also performed to better visualize the Tafel reaction mechanism (Figure 10(b) and 10(c)). The electron cloud represents both positive and negative parts of the wavefunction by red and blue color. The electronic cloud around hydrogen is highlighted by a red dotted circle. Figure 10 b represents the HOMO of the Tafel TS and the orbitals around H$_2$ formed during Tafel TS formation are highlighted by blue. This means that the orbital mixing is out of phase. The red cloud around H$_2$ in the LUMO of the Tafel TS suggests in-phase interaction of electronic wavefunctions. The phase or orbital is a direct consequence of the wave like property of electrons and generally the in-phase mixing suggests lower energy state and the out of phase mixing indicates anti-bonding orbitals or higher energy state. The corresponding TOF in both the gas and solvent phases was calculated to be $6.19 \times 10^{-45}$ sec$^{-1}$ and $1.25 \times 10^{-58}$ sec$^{-1}$, respectively. This TOF value is very low, and hence the process is less likely to take place.

It is now clear from the data (Table 6) that Tafel barrier is much higher than the calculated Heyrovsky barrier and hence Volmer-Heyrovsky reaction will be more assertive than the Volmer-Tafel reaction when using 2D monolayer Mn-MoS$_2$ material-based catalyst. Therefore, heteroatom doping in the pristine 2D monolayer MoS$_2$ has led to significant change in electronic properties of the material. As shown in our present computed results, 2D monolayer Mn-MoS$_2$ TMD material shows excellent electrocatalytic performance. The results of the descriptor-based method aided by the DFT computations have been thoroughly discussed above. This indicates that the 2D Mn-MoS$_2$ driven catalysis is a viable and efficient hydrogen production method.

**Table 6:** All reaction barriers in HER mechanism using 2D monolayer Mn-MoS$_2$ are reported here. The values of various energy changes ($\Delta G$, $\Delta E$, and $\Delta H$) are expressed in kcal mol$^{-1}$.

| Activation Barrier | $\Delta G$ (kcal mol$^{-1}$) in gas phase | $\Delta E$ (kcal mol$^{-1}$) in solvent phase | $\Delta H$ (kcal mol$^{-1}$) in solvent phase | $\Delta G$ (kcal mol$^{-1}$) in solvent phase |
|---|---|---|---|---|
| **Volmer reaction barrier** | 7.23 | 11.84 | 9.60 | 10.34 |
| **Heyrovsky reaction barrier** | 10.59 | 12.56 | 10.36 | 10.79 |
| **Tafel reaction barrier** | 90.13 | 93.66 | 92.27 | 93.72 |

# 4. Conclusion

In summary, we developed a 2D monolayer of Mn doped MoS$_2$ catalyst for HER with the aid of DFT simulations. By applying the first principles based B3LYP-D3 method, we studied the electronic properties i.e., band structure and total density of states (DOS) of the material. The DFT-D method applied to the periodic 2D slab of Mn-MoS$_2$ showed that it has zero band gap, and the DOS calculations showed that it became electron rich due to the addition of Mn in MoS$_2$. In this comprehensive study, we have encapsulated the relationship between the structure and morphology of the material that characterizes its catalytic activity. The examination of the performance of 2D monolayer Mn-MoS$_2$ TMD material for catalytic activity has been done through the Mn$_1$Mo$_9$S$_{21}$ molecular cluster model. The detailed reaction mechanism along with the transition states has been calculated by M06-L DFT method considering the finite non-periodic molecular cluster model system Mn$_1$Mo$_9$S$_{21}$. The H$_2$ evolution reaction followed two electron transfer kinetics with highly favorable Volmer-Heyrovsky mechanism. The Volmer and Heyrovsky barriers were computed to be 10.34 kcal mol$^{-1}$ and 10.79 kcal mol$^{-1}$, respectively, in the solvent phase. Lowering of the activation barrier is one of the key features of the catalyst, and the electronic overlap between the *s*-orbitals of H and the *d*-orbitals of the transition metal in TMDs has favored the H$_2$ formation during HER process. Low activation barrier energies, high TOF (7.74 ×10$^4$ sec$^{-1}$) and the theoretically determined Tafel slope (29.58 mV dec$^{-1}$) are attributed to electrochemical stability during HER, and the 2D monolayer Mn-doped MoS$_2$ TMD has become a promising and efficient electrocatalyst for HER. The strategy used in this work can also be extended to model and design other low cost and high efficiency catalysts.

## Author Contributions:

Dr. Pakhira developed the complete idea of this current research work, and he computationally studied the electronic structures and properties of the 2D Mn-MoS$_2$ TMD. Dr. Pakhira and Mr. Joy Ekka explored the whole reaction pathways; transitions states and reactions barriers and Dr. Pakhira explained the HER mechanism by the DFT calculations. Quantum calculations and theoretical models were designed and performed by Dr. Pakhira and Mr. Joy Ekka. Dr. Pakhira and Mr. Joy Ekka wrote the whole manuscript and prepared all the tables and figures in the manuscript. Prof. Keil edited the manuscript. Mr. Shrish Nath Upadhyay helped Dr. Pakhira to organize the manuscript.

## AUTHOR INFORMATION

**Corresponding Author**

**Dr. Srimanta Pakhira** − *Department of Physics, Indian Institute of Technology Indore (IIT Indore),* Khandwa Road, Simrol, *Indore, MP 453552, India.*

*Department of Metallurgy Engineering and Materials Science (MEMS), Indian Institute of Technology Indore (IIT Indore),* Khandwa Road, Simrol, *Indore, MP 453552, India.*

*Centre for Advanced Electronics (CAE), Indian Institute of Technology Indore (IIT Indore),* Khandwa Road, Simrol, *Indore, MP 453552, India.*

ORCID: orcid.org/0000-0002-2488-300X;
Email: spakhira@iiti.ac.in or spakhirafsu@gmail.com

**Authors**

**Joy Ekka** − *Department of Physics, Indian Institute of Technology Indore (IIT Indore),* Khandwa Road, Simrol, *Indore, MP 453552, India.*
ORCID: orcid.org/0000-0002-1903-6658.

**Shrish Nath Upadhyay** − *Department of Metallurgy Engineering and Materials Science (MEMS), Indian Institute of Technology Indore (IIT Indore),* Khandwa Road, Simrol, *Indore, MP 453552,* India.
ORCID: orcid.org/0000-0003-0029-4160.

**Prof. Frerich J. Keil** − Department of Chemical Reaction Engineering, Hamburg University of Technology, 21073 Hamburg, Germany.

ORCID: orcid.org/ 0000-0002-4051-6824.

## Acknowledgment:

This work was financially supported by the Science and Engineering Research Board-Department of Science and Technology (SERB-DST), Government of India under the Grant No. ECR/2018/000255. Dr. Srimanta Pakhira thanks the Science and Engineering Research Board, Department of Science and Technology (SERB-DST), Govt. of India for providing his highly prestigious Ramanujan Faculty Fellowship under the scheme no. SB/S2/RJN-067/2017, and for his Early Career Research Award (ECRA) under the grant No. ECR/2018/000255. Mr. Upadhyay thanks Indian Institute of Technology Indore, MHRD, Government of India for providing the doctoral fellowship. The authors would like to acknowledge the SERB-DST for providing the computing cluster and programs. We acknowledge National Supercomputing Mission (NSM) for providing computing resources of 'PARAM Brahma' at IISER Pune, which is implemented by C-DAC and supported by the Ministry of Electronics and Information Technology (MeitY) and Department of Science and Technology (DST), Government of India.

**Notes**

The authors declare no competing financial interest. There is no conflict of interests.

## Graphical Abstract

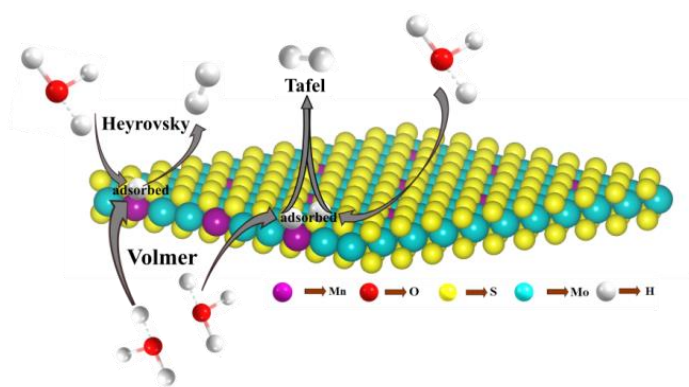